# PARALLELS OF HUMAN LANGUAGE IN THE BEHAVIOR OF BOTTLENOSE DOLPHINS

Running tittle: PARALLELS OF HUMAN LANGUAGE IN DOLPHINS

R. Ferrer-i-Cancho (1), D. Lusseau (2,3) and B. McCowan (4)

(1) Complexity and Quantitative Linguistics Lab. LARCA Research Group. Departament de Ciències de la Computació. Universitat Politècnica de Catalunya. Barcelona, Catalonia, Spain.

(2) Institute of Biological and Environmental Sciences, University of Aberdeen, Aberdeen, UK

(3) Technical University of Denmark, Lyngby, Denmark

(4) Population Health and Reproduction, School of Veterinary Medicine, University of California, Davis, CA, USA

**ABSTRACT:** Dolphins exhibit striking similarities with humans. Here we review them with the help of quantitative linguistics and information theory.

**KEYWORDS:** linguistic laws; quantitative linguistics; information theory; dolphins

## WHY BOTTLENOSE DOLPHINS?

An important reason to investigate dolphins is that they exhibit striking similarities with humans. Like us, they use tools: dolphins break off sponges and wear them over their rostrum while foraging on the seafloor (Smolker et al 1997). Dolphins are also capable of recognizing their body in front of a mirror (Reiss, Marino 2001). Closely related with their capacity to see through sound is their capacity to form abstract representations that are independent from modality (Herman et al 1998). Dolphins share with us other traits that are appealing from the perspective of language theory. First, they exhibit spontaneous vocal mimicry (Reiss, McCowan 1993) which suggests a predisposition to learn a vocal communication system. Second, they live, in general, in fission-fusion societies and display complex social behaviours (Lusseau et al 2003; Connor, Krützen 2015) while converging research supports that the complexity of a society and the complexity of communication are correlated (Freeberg et al 2012). Third, they can learn a signal to innovate, namely to show a behavior not seen in the current interaction session (Foer 2015). This tells us something about the limits on memory and creativity in dolphins and is challenging from a theoretical perspective:

many researchers believe that a crucial difference between humans and other species is our unbounded capacity to generate sequences, e.g., by embedding sentences into other sentences (e.g., Gregg 2013; Hauser et al 2002), or a capacity for large lexicons (Hurford 2004). In short, bottlenose dolphins share many traits we associate as pre-requisite for our complex linguistic abilities.

Although possessing such an infinite capacity makes a qualitative difference compared to a species with a finite capacity, the fact is that (a) a species being able to generate a huge number of sentences may not be distinguishable from a species that has infinite capacity (supposing that the latter is really true) and (b), humans have problems with parsing sentences with just a few levels of embedding (Christiansen, Chater 2015). The point is that the problem of infinite versus finite capacity does seem to be well poised and that dolphin capacity to innovate is being overlooked. We humans are fascinated by infinity (perhaps for purely aesthetical reasons) and may have rushed to steal the flag of infinity to keep it in some anthropocentric fortress where other species are not allowed to get in.

In a recent book, the parallel in cognitive abilities between humans and dolphins has been questioned (Gregg 2013). On p. 158, its author compares dolphins against chickens, chimpanzees, bees and humans by means of various attributes. For instance, humans get a score of 5 for the attribute "limitless expression" while bees get a 1 and the other species get a zero. Concerning the attribute "recursion", humans get a 5, dolphins and chimpanzees get a 1, and chickens and bees get a zero. As far as we know, there is no replicable scientific approach to determine whether a source has "limitless expression" or "recursion". Furthermore, it can be argued that the very notion of recursion, borrowed from the generativist tradition in linguistics (e.g., Hauser et al 2002), has a definition that is too fuzzy or weak to allow for an operational definition that can be used for real testing. In this context, it is suspicious that humans always get the maximum scores in Gregg's (2013) table. The statistical analysis is too poor (scores are based on ill-defined qualitative units and statistical testing or model selection are missing). The approach itself is not objective at all, and is an illustration of an anthropocentric bias interfering scientific work.

## TOWARDS A LESS BIASED PERSPECTIVE

*A priori*, there are at least three approaches in a comparative research. First, one that puts humans at the center: what humans have and other species do not (or we do not know if they do...) as we have already seen. Second, one that puts dolphins at the center, what dolphins have and other species (including humans) do not. For instance, we could score humans and other species by their swimming or sonar skills (that would easily give a high score to dolphins and a low score to humans, chickens, bees, and chimpanzees), their capacity to understand the innovate signal (it would be great to have comparison of the capacity across species to understand this signal), and so on. A problem of the two approaches above is that they use a species as the reference point. The solution to the anthropocentric bias is not a dolphin bias. This leads to a third approach, one that is a priori neutral, resembling the eye of a physicist looking at the universe with humility.

The neutral approach has to be rooted on the scientific method (one should get rid of tricks to make a species win and poor statistical analyses). We suggest that the solution has to combine two perspectives (not exclusively). First, the perspective of quantitative linguistics, which is specifically concerned about measurement (vs qualitative scoring), counting, statistical testing and thus offers a more neutral approach for analysis and comparison. Second, the perspective of information theory and physics as the theoretical framework: some level of abstraction is needed to be able to see some unity across species or levels of organization of life. Without abstraction, fragmentation of knowledge is inevitable but not necessarily and intrinsic property of reality.

One of the purposes of quantitative linguistics is the study of statistical laws of language. Perhaps the most popular law of language is Zipf's law for word frequencies, which defines the frequency of a word as a function of its rank (the most frequent word has rank 1, the 2nd most frequent word has rank 2 and so on). This law is an empirical law that was popularized (but not discovered by Zipf) and states that the frequency of *i*-th most frequent word of a text follows *approximately* (Zipf 1949)

$$f \propto i^{-\alpha}, \tag{1}$$

with $\alpha \approx 1$. For many decades, the law has been considered to be universal, i.e. in the sense of holding approximately in every language where it has been tested but large-scale analyses have unveiled deviations from Eq. 1 across languages (Bentz et al 2015; Mehri, Jamaati 2017; Yu et al 2018). Interestingly, patterning consistent with the law has been found in dolphin vocalizations (Markov, Ostrovskaya, 1990, McCowan et al 1999), suggesting at least some common communicative background between dolphins and certain languages.

Other examples of laws stemming from Zipf’s (1935, 1949) foundational work are the law of meaning distribution, i.e. the tendency of more frequent words to have more meanings, and the law of abbreviation, i.e. the tendency of more frequent words to be shorter. Another law stemming from the seminal work of P. Menzerath’s (1954) and G. Altmann (1980) is Menzerath’s law, which indicates that the longer the linguistic construct, the shorter its parts. Interestingly, Zipf’s law for word frequencies (McCowan et al 1999), the law of meaning distribution (Ferrer-i-Cancho, McCowan 2009), the law of abbreviation (Ferrer-i-Cancho, Lusseau 2009, Vradi 2021) and Menzerath’s law (Vradi 2021) have been found in the behavior of dolphins. The law of meaning distribution has also been found in chimpanzee gestures (Hobaiter and Byrne 2014). The law of abbreviation has also been found in other species (Semple et al 2010; Luo et al 2013) and as well as in the genetic code (Naranan, Balasubrahmanyan 2000). Menzerath’s law also holds in music (Boroda, Altmann 1991), the vocalizations of geladas (Gustison et al 2016) and macromolecoules (Wilde, Schwibbe 1989; Ferrer-i-Cancho, Forns 2010; Li 2012, Nikolaou 2014; Shahzad et al 2015). For a recent review of these laws in biology, see Semple et al (2022).

The motivation of research on laws of language beyond human language is three-fold. First is the need to check if there are true, cross-species, universals; properties that hold in large ensembles of species, extending the field of the typology of linguistic universals to animal behavior and beyond (e.g., the genetic code). Second is the need of general principles of organization of behavior and living systems that the ubiquity of those patterns suggests. There may not be truly universal statistical laws but there might be universal principles that can help us to understand not only patterns (laws) but also deviations from them or even their disappearance. Third is providing new tools for comparative psychology to answer questions such as: what is the closest species to ours? Fourth is breaking the illusion of autonomy of certain fields such as linguistics, animal behavior, biology, physics and computer science: it is not only possible but critical to unify them to better understand the role that information plays in living systems.

QUANTITATIVE LINGUISTICS OF DOLPHIN BEHAVIOR

Dolphins produce whistles, narrow band vocalizations that are rather easy to investigate compared to other kinds of vocalizations. The finding of patterning consistent with Zipf’s law for word frequencies in dolphin whistles (McCowan et al 1999) and other dolphin vocalizations (Markov, Ostrovskaya 1990) is very suggestive: dolphins and we humans could be sharing similar principles of communication (Ferrer-i-Cancho, Solé 2003; Ferrer-i-Cancho 2016). Zipf thought that the law revealed principles of the organization of vocabularies, in particular a conflict between speaker and hearer needs and in general a conflict between unification and diversification forces (Zipf 1949). For this reason, Zipf can be considered as a precursor of modern complexity theory, which regards complexity as emerging in a critical balance between order and disorder (Kauffman 1993). Recently, Zipf’s view has been formalized mathematically using information theory, leading to the hypothesis that Zipf's law is a local optimum of conflicting communication principles (Ferrer-i-Cancho, Diaz-Guilera 2007).

The relevance of Zipf’s law in human language has been questioned with the argument that random typing, i.e. a random sequence of characters such as

> *wbqcrw h q rorjleabeyxkrlpqkpchnesguliwkb mrltn q a rss vfs w a h rlzpxxtxbkqetfwfpqudgwaorqwgqmo wyngwtbseuodboxaw x rldua eucx mmard xgqzv uu pueuerc pkizuauyrwi bllhjddv bp anud xbxvjyymioymvzebc tdtsecdijntssyepqdubcvxjd evavybwvejp w z uvspufvdvuzyf t nllifznwatic* reproduces Zipf’s law for word frequencies (Mandelbrot 1953; Miller 1957; Miller, Chomsky 1963).

A similar argument is Suzuki et al’s (2005) die rolling experiment where characters are replaced by the sides of a die and one especial side (6 in the example below) plays the role of a space (a word delimiter). For instance, the sequence of rolls

1221452**6**215152151**6**53**66**2**6**434**6**2343**6**54**66**

gives the sequence of pseudowords (_ is used to indicate empty words)

1221452, 215152151, 53, _ , 2, 434,2343, 54, _.

Interestingly, the proponents of these models have never combined in a plot the actual rank histogram these models produce and the real shape of Zipf’s law in human language. Even more worrying is the fact that they have never performed a statistically rigorous test of the adequacy of random typing. When a careful statistical comparison has been made, radical differences between random typing and real language have surfaced (Ferrer-i-Cancho, Elvevåg 2009; see also Ferrer-i-Cancho, Gavaldà 2009).

The random typing hypothesis for human language is easy to dismantle looking at the rank spectrum (Ferrer-i-Cancho, Elvevåg 2009) or the frequency spectrum (Ferrer-i-Cancho, Gavaldà 2009) but also looking at the statistical properties of the sequence of 'words' produced. While random typing (Miller 1957; Miller, Chomsky 1963) and die rolling (Suzuki et al 2005; Niyogi, Berwick 1995) produce a sequence of independent 'words', long range correlations characterize real texts when regarded as sequences of words (Montemurro, Zanette 2011; Montemurro, Pury 2002) or sequences of letters (Moscoso del Prado Martín 2011; Ebeling, Pöschel 1994). For similar reasons, it can be concluded that dolphins do not roll dies although some researchers leave open the possibility to explain Zipf’s law in dolphins’ whistles (Suzuki et al 2005). Consider a sequence of dolphin whistles produced by Panama in McCowan et al’s (1999) dataset:

25, 4, 3, 3, 1, 27, 12.

Whistle type 25 is at distance 1 of whistle type 4, and distance 4 of whistle type 1. The analysis of the correlation between whistles at a certain distance in 17 dolphins from McCowan et al’s dataset (1999) shows that significant correlations between whistles at a certain distance are found and that they are reliable up to distance 4 (Ferrer-i-Cancho, McCowan 2012). Indeed, rather long-range correlations are found across species (Kershenbaum et al 2014). These statistical dependencies between elements of the sequence are incompatible with die rolling.

The law of meaning distribution dictates that more frequent words tend to have more meanings (Zipf 1949). The contexts of use of a word can be regarded as a proxy for their meaning, and the same for the context of use of a call in another species. Interestingly, dolphin whistles show a tendency of more frequent whistles to be used in more behavioral contexts (play, sex, food,...) (Ferrer-i-Cancho, McCowan 2009). These correlations are at the core of models of Zipfian laws that consider word frequency as an epiphenomenon of the number of meanings of a word (Ferrer-i-Cancho, Díaz-Guilera 2007; Ferrer-i-Cancho 2018, Ferrer-i-Cancho & Vitevitch 2018). For instance, a simple information theoretic model of Zipf’s law for word frequencies assumes that the probability of a word is proportional to its number of meanings (Ferrer-i-Cancho 2005).

The law abbreviation, i.e. the tendency of more frequent words to be shorter (Zipf 1935), holds practically in every language where it has been tested (Bentz, Ferrer-i-Cancho 2015) and it is a robust pattern, i.e. it holds regardless of the magnitude: letters (Bent, Ferrer-i-Cancho 2015), duration in time (Hernández-Fernández et al 2019), number of strokes in Japanese kanjis or Chinese characters (Sanada 2008; Wang, Chen 2015). Thus, the law can be generalized as a tendency of more frequency types to have a greater magnitude. The law was hypothesized to originate from the minimization of a mean cost of words by Zipf himself (Zipf 1949) and his view can be regarded as a precursor of the problem of compression in standard information theory (Cover, Thomas 2006). Recently, an intimate relationship between the minimization of the mean cost of types and the law of abbreviation has been shown (Ferrer-i-Cancho et al 2020; Ferrer-i-Cancho et al 2013).

The law of abbreviation has been found in other species. The first example is the pioneering research in chick-a-dee calls (Hailman et al 1987; Hailman et al 1985; Ficken et al 1978). Other examples are the vocalizations of Formosan macaques (Semple et al 2010) or bats (Luo et al 2013). In the context of non-vocal behavior, the law of abbreviation has also been found in sign language (Börstell, C. et al 2016) and in a subset of chimpanzee gestures (Heesen et al 2019). Interestingly, the generalized version of the law was found in the behavior of dolphins living in the fiords of New Zealand (Ferrer-i-Cancho, Lusseau 2009). The surface behavior of dolphins consists of about 30 patterns that are composed of elementary behavioral units, e.g.

side flop: jump + side

tail-slap: two + hit + tail

spy hop: stationary + expose + head

tail-stock dive: arch

Accordingly, tail-stock dive has size 1, side flop has size 2, tail-slap and spy hope has size 3. The generalized law of abbreviation is found as a tendency of the number of elementary behavioral units to decrease as the frequency of the behavioral pattern increases. Recently, it has been shown that Zipf's law of abbreviation also holds in dolphin vocal behavior: the duration of more frequent whistle types tends to be shorter (Vradi 2021).

Rather long-range correlations are found in the sequences of dolphin surface behavioral patterns (Ferrer-i-Cancho, Lusseau 2006) and in dolphin whistle sequences (Ferrer-i-Cancho, McCowan 2012), as it is also the case of the behavior of other species (Kershenbaum et al 2014). This kind of correlations in the sequence of behaviors are appealing because they challenge the view that they are unique to active communication in humans (Ferrer-i-Cancho et al 2008).

Recently, it has been shown that dolphin whistle sequences resemble human language sequences for exhibiting patterning consistent with Menzerath's law: sequences formed by a larger number of whistles tend to be formed of shorter whistles (Vradi 2021). All these findings together strongly suggest that dolphin whistles exhibit some form of lexical syntax, namely structured sequences where units are meaningful, as opposed to phonological syntax, where units lack meaning (Marler 1998). Prototypical examples of each kind of syntax are human language and bird song, respectively.

## TOWARDS THE FUTURE

Readers interested in the origins of these laws from a theoretical perspective have two major options. One the one hand, they can buy the traditional collections of random models (e.g., die rolling) that are focused on modelling one pattern individually, regardless of other patterns or related phenomena, with no concern about plausibility, jeopardizing the construction of a parsimonious theory of language. One the other hand, they can opt to consider models that attempt to provide explanations in a unified and compact fashion, defining what could be generously called a prototheory at this moment. The first option sustains by itself because (1) scientific knowledge relies mostly on trust and maximum credit is assigned to certain individuals and research institutions; (2) the dynamics of citations is self-reinforcing; and (3) the illusion of parsimony caused by models of local simplicity as opposed to general or more ambitious prototheories.

For readers interested in the second option, we suggest that they take look at information theoretic models that shed light on the origins of Zipf's law for word frequencies (Ferrer-i-Cancho 2018; Ferrer-i-Cancho 2016), Zipf's law of abbreviation (Ferrer-i-Cancho et al 2020; Ferrer-i-Cancho et al 2013), and Menzerath's law (Gustison et al 2016; Ferrer-i-Cancho et al 2020), the principle of contrast and a vocabulary learning bias in children (Ferrer-i-Cancho 2016; Carrera-Casado, Ferrer-i-Cancho 2021). These models can be locally heavier than toy models such as random typing, but have the potential for a more compact theory of how natural systems are and how they evolve.

In this short piece, we have reviewed many statistical similarities between human language and dolphin behavior. This comparison is still in its infancy and other researchers are invited to contribute with fresh ideas, new data and updated methods. Explorations of "linguistic" laws that have only been investigated in humans can help us to establish new connections across species and disciplines. Research on other species is confirming the presence of "linguistic" laws in many other species (see Semple et al 2022 for a review).

Inspired by Zipf's seminal's work (1935, 1949), a unified information theoretic framework is available to explain some of those similarities as the outcome of optimized behavior (Gustison et al 2016, Ferrer-i-Cancho et al 2020, Carrera-Casado, Ferrer-i-Cancho 2021). We reiterate that the origin of the similarities between human language and dolphin behavior does not need to be linguistic or communicative (e.g., compression can be reduced to a principle of cost minimization). While the reason may not be due to a sophisticated cognition, these regularities and their underlying mechanisms can make us more aware of the common ground that we share with other species and the potential for unifying views across disciplines.

ACKNOWLEDGEMENTS

RFC is grateful to A. Mandich and her collaborators for the opportunity to participate in the International Meeting on Consciousness & Communication in dolphins (Genoa, 5 May 2015). RFC was funded by the grants 2014SGR 890 (MACDA) from AGAUR (Generalitat de Catalunya) and also the APCOM project (TIN2014-57226-P) from MINECO (Ministerio de Economia y Competitividad).

REFERENCES

Altmann, G., 1980. Prolegomena to Menzerath's law. *Glottometrika,* 2, 1–10.

Bentz, C., Ferrer-i-Cancho, R., 2015. Zipf's law of abbreviation as a language universal. Capturing Phylogenetic Algorithms for Linguistics. Lorentz Center Workshop, Leiden, October 2015

Bentz, C., Verkerk, A., Kiela, D., Hill, F., Buttery, P., 2015. Adaptive communication: languages with more non-native speakers tend to have fewer word forms. *PLoS ONE,* 10 (6), e0128254.

Boroda, M.G., Altmann, G., 1991. Menzerath's law in musical texts. *Musikometrika,* 3, 1–13.

Börstell, C., Hörberg, T., & Östling, R., 2016. Distribution and duration of signs and parts of speech in Swedish Sign Language. *Sign Language & Linguistics*, 19(2), 143–196.

Carrera-Casado, D., Ferrer-i-Cancho, R., 2021. The advent and fall of a vocabulary learning bias from communicative efficiency. *Biosemiotics,* 14, 345-375.

Christiansen, M. H., Chater, N., 2015. The language faculty that wasn't: a usage-based account of natural language recursion. *Frontiers in Psychology,* 6, 1182.

Connor, R. C., Krützen, M., 2015. Male dolphin alliances in Shark Bay: changing perspectives in a 30-year study. *Animal Behaviour,* 103, 223–235.

Cover, T. M., Thomas, J. A., 2006. *Elements of information theory*, 2nd edition. Hoboken, NJ: Wiley.

Ebeling, W., Pöschel, T., 1994. Entropy and long-range correlations in literary English. *Europhysics Letters,* 26 (4), 241-246.

Ferrer-i-Cancho, R., 2005. Zipf's law from a communicative phase transition. *European Physical Journal B,* 47, 449-457.

Ferrer-i-Cancho, R., 2016. The optimality of attaching unlinked labels to unlinked meanings. *Glottometrics,* 36, 1-16.

Ferrer-i-Cancho, R., 2018. Optimization models of natural communication. *Journal of Quantitative Linguistics,* 25 (3), 207-237.

Ferrer-i-Cancho, R., 2016. Compression and the origins of Zipf’s law for word frequencies. *Complexity,* 21, 409-411.

Ferrer-i-Cancho, R., Bentz, C, Seguin, C., 2020. Optimal coding and the origins of Zipfian laws. *Journal of Quantitative Linguistics*, in press. doi: 10.1080/09296174.2020.1778387

Ferrer-i-Cancho, R., Díaz-Guilera, A., 2007. The global minima of the communicative energy of natural communication systems. *Journal of Statistical Mechanics*, P06009.

Ferrer-i-Cancho, R., Elvevåg, B., 2009. Random texts do not exhibit the real Zipf's-law-like rank distribution. *PLoS ONE,* 5(4), e9411.

Ferrer-i-Cancho, R., Forns, N., 2010. The self-organization of genomes. *Complexity,* 15 (5), 34–36.

Ferrer-i-Cancho, R., Gavaldà, R., 2009. The frequency spectrum of finite samples from the intermittent silence process. *Journal of the American Association for Information Science and Technology,* 60(4), 837-843.

Ferrer-i-Cancho, R., Hernández-Fernández, A., Lusseau, D., Agoramoorthy, G., Hsu, M. J., Semple, S., 2013. Compression as a universal principle of animal behavior. *Cognitive Science,* 37(8), 565-1578.

Ferrer-i-Cancho, R., Lorenzo, G., Longa, V., 2008. Long-distance dependencies are not uniquely human. In: *The Evolution of Language: Proceedings of the 7th International Conference (EVOLANG7)*. Smith, A. D. M., Smith, K., Ferrer i Cancho, R (eds.). Singapore: World Scientific Press, pp. 115-122.

Ferrer-i-Cancho, R., Lusseau, D., 2006. Long-term correlations in the surface behavior of dolphins. *Europhysics Letters,* 74(6), 1095-1101.

Ferrer-i-Cancho, R., Lusseau, D., 2009. Efficient coding in dolphin surface behavioral patterns. *Complexity,* 14(5), 2325.

Ferrer-i-Cancho, R., McCowan, B., 2009. A law of word meaning in dolphin whistle types. *Entropy,* 11(4), 688701.

Ferrer-i-Cancho, R., McCowan, B., 2012. The span of dependencies in dolphin whistle sequences. *Journal of Statistical Mechanics*, P06002.

Ferrer-i-Cancho, R. & Vitevitch, M. S., 2018. The origins of Zipf's meaning-frequency law. *Journal of the American Society for Information Science and Technology*, 69(11), 1369-1379.

Ficken, M. S., Hailman, J. P., Ficken, R. W., 1978. A model of repetitive behaviour illustrated by chickadee calling. *Animal Behaviour* 26(2), 630-631.

Foer, J., 2015. It’s time for a conversation. Breaking the communication barrier between dolphins and humans. http://ngm.nationalgeographic.com/2015/05/dolphin-intelligence/foer-text#

Freeberg, T. M., Dunbar, R.I. M., Ord, T. J., 2012. Social complexity as a proximate and ultimate factor in communicative complexity. *Philosophical Transactions of the Royal Society B,* 367, 1785-1801.

Gregg, J., 2013. *Are dolphins really smart? The mammal behind the myth*. Oxford University Press, Oxford.

Gustison, M.L, Semple, S., Ferrer-i-Cancho, R., Bergman, T. J., 2016. Gelada vocal sequences follow Menzerath’s linguistic law. *Proceedings of the National Academy of Sciences USA,* 113(19), E2750-E2758.

Heesen, R, Hobaiter, C., Ferrer-i-Cancho, R., Semple, S., 2019. Linguistic laws in chimpanzee gestural communication. *Proceedings of the Royal Society B: Biological Sciences*, 20182900.

Kauffman, S. A., 1993. The origins of order: self-organization and selection in evolution. New York: Oxford University Press.

Kershenbaum A., Bowles A.E., Freeberg T.M., Jin, D.Z., Lameira A.R., Bohn, K., 2014. Animal vocal sequences: not the Markov chains we thought they were. *Proceedings of the Royal Society B,* 281(1792), 20141370.

Hailman, J. P., Ficken, M. S., Ficken, R. W., 1985. The 'chick-a-dee' calls of Parus atricapillus: a recombinant system of animal communication compared with written English. *Semiotica*, 56, 191-224.

Hailman, J. P., Ficken, M. S., Ficken, R., 1987. Constraints on the structure of combinatorial "chick-a-dee" calls. *Ethology* 75, 62-80.

Hauser, M.D., Chomsky, N., Fitch, W. T., 2002. The faculty of language: what is it, who has it, and how did it evolve? *Science,* 298(5598), 1569-1579.

Herman, L. M., Pack, A. A., Homann-Kuhnt, M., 1998. Seeing through sound: dolphins perceive the spatial structure of objects through echolocation. *Journal of Comparative Psychology,* 112, 292-305.

Hernández-Fernández, A., G. Torre, I., Garrido, J.M., Lacasa, L., 2019. Linguistic laws in speech: the case of Catalan and Spanish. *Entropy*, 21(12), 1153.

Hobaiter, C., Byrne, R. W, 2014. The meanings of chimpanzee gestures. *Current Biology*, 24(14), 1596-1600.

Hurford, J.R., 2004. Human uniqueness, learned symbols and recursive thought. *European Review,* 12(4), 551-565.

Li, W., 2012. Menzerath's law at the gene-exon level in the human genome. *Complexity* 17, 49-53.

Luo, B., Jiang, T., Liu, Y., Wang, J., Lin, A., Wei, X., Feng, J., 2013. Brevity is prevalent in bat short-range communication. *Journal of Comparative Physiology A,* 199(4), 325-333.

Lusseau, D., Schneider, K., Boisseau, O. J., Haase, P., Slooten, E., Dawson, S. M., 2003. The bottlenose dolphin community of doubtful sound features a large proportion of long-lasting associations - can geographic isolation explain this unique trait? *Behavioral Ecology and Sociobiology,* 54(4), 396-405.

Mandelbrot, B., 1953. An informational theory of the statistical structure of language. In Jackson, W. (ed.), Communication theory. London: Butterworths, pp. 486-502.

Markov, V.I. and Ostrovskaya, V.M., 1990. Organization of communication system in *Tursiops truncatus montagu*. In *Sensory abilities of cetaceans* (Thomas, J.A. and Kastelein, R.A. et al., eds), pp. 599–622, Springer

Marler, P., 1998. Animal communication and human language. In Jablonski, G., and Aiello, L.C. (Eds.), The origin and diversification of language. Wattis Symposium

Series in Anthropology. Memoirs of the California Academy of Sciences 24. ¡San Francisco, CA: California Academy of Sciences, pp. 1-19.

McCowan, B., Hanser, S. F., Doyle, L. R., 1999. Quantitative tools for comparing animal communication systems: information theory applied to bottlenose dolphin whistle repertoires. *Animal Behaviour* 57, 409-419.

Menzerath, P., 1954. Die Architektonik des deutschen Wortschatzes. Dümmler: Bonn.

Mehri, A., Jamaati, M., 2017. Variation of Zipf's exponent in one hundred live languages: a study of the Holy Bible translations. *Physics Letters, A* 381, 2470–2477.

Miller, G. A., 1957. Some effects of intermittent silence. *American Journal of Psychology* 70, 311-314.

Miller, G. A., Chomsky, N., 1963. Finitary models of language users. In: Luce, R. D., Bush, R., Galanter, E., (eds.), *Handbook of Mathematical Psychology*, volume 2, pages 419-491. New York: Wiley.

Montemurro, M., Pury, P. A., 2002. Long-range fractal correlations in literary corpora. *Fractals*, 10, 451-461.

Montemurro, M. A., Zanette, D. H., 2011. Universal entropy of word ordering across linguistic families. *PLoS ONE,* 6, e19875.

Moscoso del Prado Martín, F., 2011. The universal 'shape' of human languages: spectral analysis beyond speech. Available from Nature Precedings
http://hdl.handle.net/10101/npre.2011.6097.1

Naranan, S., Balasubrahmanyan, V.K., 2000. Information theory and algorithmic complexity: applications to linguistic discourses and DNA sequences as complex systems. Part I: Efficiency of the genetic code of DNA. *Journal of Quantitative Linguistics,* 7 (2), 129-151.

Nikolaou, C., 2014. Menzerath-Altmann law in mammalian exons reflects the dynamics of gene structure evolution. *Computational Biology and Chemistry* 53, 134–143.

Niyogi, P., Berwick, R. C., 1995. A note on Zipf's law, natural languages, and noncoding DNA regions. A.I. Memo No. 1530 / C.B.C.L. Paper No. 118.

Reiss, D., Marino, L., 2001. Mirror self-recognition in the bottlenose dolphin: a case of cognitive convergence. *Proceedings of the National Academy of Sciences,* 98(10), 5937-5942.

Reiss, D., McCowan, B., 1993. Spontaneous vocal mimicry and production by bottlenose dolphins (tursiops truncatus): evidence for vocal learning. *Journal of Comparative Psychology,* 107(3), 301-312.

Semple, S., Ferrer-i-Cancho, R., Gustison, M., 2022. Linguistic laws in biology. *Trends in Ecology and Evolution* 37(1), 53-66.

Semple, S., Hsu, M. J., Agoramoorthy, G., 2010. Efficiency of coding in macaque vocal communication. *Biology Letters,* 6, 469-471.

Shahzad, K., Mittenthal, J.E., Caetano-Anollés, G., 2015. The organization of domains in proteins obeys Menzerath-Altmann's law of language. *BMC Systems Biology* 9(44), 1-13.

Smolker, R., Richards, A., Connor, R., Mann, J., Berggren, P., 1997. Sponge carrying by dolphins (delphinidae, tursiops sp.): A foraging specialization involving tool use? *Ethology,* 103(6), 454-465.

Sanada, H.,2008. *Investigations in Japanese historical lexicology* (Rev. ed). Peust & Gutschmidt.

Suzuki, R., Tyack, P. L., Buck, J., 2005. The use of Zipf's law in animal communication analysis. *Animal Behaviour,* 69, F9-F17.

Vradi, A., 2021. Dolphin communication: a quantitative linguistics approach. Master thesis, Universitat Politècnica de Catalunya, Barcelona, Catalonia (Spain) http://hdl.handle.net/2117/348201

Wang, Y., Chen, X.,2015. Structural complexity of simplified Chinese characters. In A. Tuzzi, M. Benesová, & J. Macutek (Eds.), *Recent Contributions to Quantitative Linguistics*. De Gruter.

Wilde, J., Schwibbe, M.H., 1989. Organizationsformen von Erbinformation Im Hinblick auf die Menzerathsche Regel. Das Menzerathsche Gesetz in Informationsverarbeitenden Systemen, eds Altmann, G., Schwibbe, M.H., Kaumanns, W. (G. Olms, Hildesheim), pp 92–107.

Yu, S., Xu, C., Liu, H., 2018. Zipf’s law in 50 languages: its structural pattern, linguistic interpretation, and cognitive motivation. http://arxiv-export-lb.library.cornell.edu/abs/1807.01855v1.

Zipf, G. K., 1935. *The psychobiology of language: an introduction to dynamic philology.* Cambridge, Mass.: M.I.T. Press.

Zipf, G. K., 1949. *Human behaviour and the principle of least effort*. Addison-Wesley, Cambridge (MA), USA.